\documentclass[
superscriptaddress,
amsmath,amssymb,
aps,
prx,
nofootinbib,
]{revtex4-2}

\usepackage{comment}
\usepackage{tabularx}
\usepackage{graphicx}
\usepackage{dcolumn}
\usepackage{bm}
\usepackage{mathtools}
\usepackage{color}
\usepackage{braket}

\usepackage{here}

\begin{document}

\title{Conveyor-belt magneto-optical trapping of molecules}

\author{Grace K. Li}
\email{kehui\_li@g.harvard.edu}
\author{Christian Hallas}
\author{John M. Doyle}
\affiliation{Department of Physics, Harvard University, Cambridge, MA 02138, USA}
\affiliation{Harvard-MIT Center for Ultracold Atoms, Cambridge, MA 02138, USA}

\date{\today}

\begin{abstract}
Laser cooling is used to produce ultracold atoms and molecules for quantum science and precision measurement applications. Molecules are more challenging to cool than atoms due to their vibrational and rotational internal degrees of freedom. Molecular rotations lead to the use of type-II transitions (\(F \geq F'\)) for magneto-optical trapping (MOT). When typical red detuned light frequencies are applied to these transitions, sub-Doppler heating is induced, resulting in higher temperatures and larger molecular cloud sizes than realized with the type-I MOTs most often used with atoms. To improve type-II MOTs, Jarvis et al. \cite{RbBlueMOT} proposed a blue-detuned MOT to be applied after initial cooling and capture with a red-detuned MOT. This was successfully implemented \cite{YOBlueMOT, SrFBlueMOT, CaFBlueMOT}, realizing colder and denser molecular samples. Very recently, Hallas et al. \cite{CaOHBlueMOT} demonstrated a blue-detuned MOT with a ``1+2" configuration that resulted in even stronger compression of the molecular cloud. Here, we describe and characterize theoretically the conveyor-belt mechanism that underlies this observed enhanced compression. We perform numerical simulations of the conveyor-belt mechanism using both stochastic Schrödinger equation (SSE) and optical Bloch equation (OBE) approaches. We investigate the conveyor-belt MOT characteristics in relation to laser parameters, $g$-factors, and the structure of the molecular system.
\end{abstract}

\maketitle

\section{Introduction}
Since their first debut in the 1980s, laser-cooling and magneto-optical trapping technologies have flourished and become cornerstones of ultracold physics and quantum science. The ability to produce dense, cold atomic samples enabled further phase-space compression, granting access to the production of quantum degenerate gases and single quantum state control within optical tweezers and optical lattices. These achievements led to advances in quantum computing \cite{bluvstein2024logical,madjarov2020high,isenhower2010demonstration}, quantum simulation \cite{labuhn2016tunable,bernien2017probing,bakr2009quantum,parsons2016site} and atomic clocks \cite{bothwell2022resolving}.
Recently, researchers have identified ultracold molecules as powerful tools for all of these areas, spurring rapid growth in molecular cooling methods. Ultracold molecules have been produced by assembling ultracold atoms via magneto-association \cite{kohler2006production,ni2008high}, photo-association \cite{aikawa2010coherent,takekoshi2014ultracold,molony2014creation}, optoelectrical Sisyphus cooling \cite{zeppenfeld2012sisyphus}, and by buffer gas cooling followed by direct laser cooling \cite{barry2014magneto,anderegg2017radio,collopy20183d,vilas2022magneto,lasner2024magneto}. Quantum computation, quantum simulation and precision searches for physics beyond the Standard Model have been demonstrated, utilizing optical traps \cite{anderegg2023quantum} and optical tweezer arrays of ultracold molecules \cite{bao2023dipolar,holland2023demand, vilas2024optical, zhang2022optical}.

\medskip
Although ultracold molecules have tremendous scientific promise, the vibrational and rotational degrees of freedom present in molecules pose challenges for laser cooling. Spontaneous radiative decay into vibrational states lead to $\sim$ 3 -- 10 repump lasers being used in order to prevent decays into unaddressed vibrational dark states. Once the system is engineered to allow for $>10,000$ photon scatters (the ``photon budget"), molecules can be efficiently cooled into the $\mu$K temperature regime and loaded into optical traps. In order to prevent loss to dark rotational states, the ground state used in molecular laser cooling has rotational quantum number $N = 1$, and the excited state is chosen to have a lower rotational angular momentum\footnote{$N'=0$ or $N'=1$, depending on the electronic angular momentum in the specific excited state.}. Transitions like this with total angular momentum $F \geq F'$ are called type-II transitions, and are used in all molecular laser cooling experiments, whereas the optical transitions in atomic laser-cooling are usually chosen to be type-I ($F < F'$) for its simplicity and fast scattering rate. For both atoms and molecules,  conventional MOTs use red-detuned lasers to produce Doppler-cooling. The distinct structures of type-I and type-II systems lead to different sub-Doppler behaviors: while a type-I system exhibits both Doppler and Sisyphus cooling, a type-II system experiences Doppler cooling and Sisyphus heating, which leads to higher temperatures and lower densities. To address these shortcomings, a blue-detuned MOT\footnote{The blue-detuned MOT is implemented after a red-detuned MOT, as the higher capture velocity of the red-detuned MOT is needed to trap a sufficient number of molecules from a molecular beam.} was proposed \cite{RbBlueMOT} and then achieved in several molecular laser cooling experiments \cite{SrFBlueMOT,YOBlueMOT,CaFBlueMOT}. In our previous work with CaOH \cite{CaOHBlueMOT}, a ``1+2" MOT configuration was proposed and realized. This approach provided exceptionally high compression of the molecular cloud, greater than the originally proposed blue MOT. In that CaOH work, the essential mechanism for this MOT behavior was identified but not fully explored.

\medskip
Here, we develop a simple and quantitatively descriptive mechanism underlying this 1+2 configuration---the conveyor-belt MOT (or conveyor MOT, for short). We theoretically characterize the performance of the conveyor MOT as a function of laser power, detunings and other parameters, and explore the effectiveness of the conveyor-belt trapping mechanism in more complex molecular systems. Finally, we propose a far-detuned conveyor MOT that has the potential to achieve high MOT capture velocities.

\section{The Conveyor-Belt Trapping Mechanism}
\label{mechanism}
It is understood that magneto-optical trapping is generated by mechanisms dependent on a non-zero excited-state $g$-factor \cite{tarbutt2015magneto}. However, recent theoretical predictions and experimental realizations have shown that magneto-optical trapping for molecules with near-zero excited-state $g$-factors \cite{tarbuttOBE1, RbBlueMOT, SrFBlueMOT, CaFBlueMOT} is also viable. The mechanism by which a trapping force can arise solely from $g$-factors in the ground states remains to be fully explained. In this section, we propose a MOT light configuration that provides robust trapping in type-II atomic and molecular systems with non-zero ground state $g$-factors, and we provide a simple, intuitive explanation for the underlying trapping mechanism.

\begin{figure}[ht]
    \centering
    \includegraphics[width=0.95\linewidth]{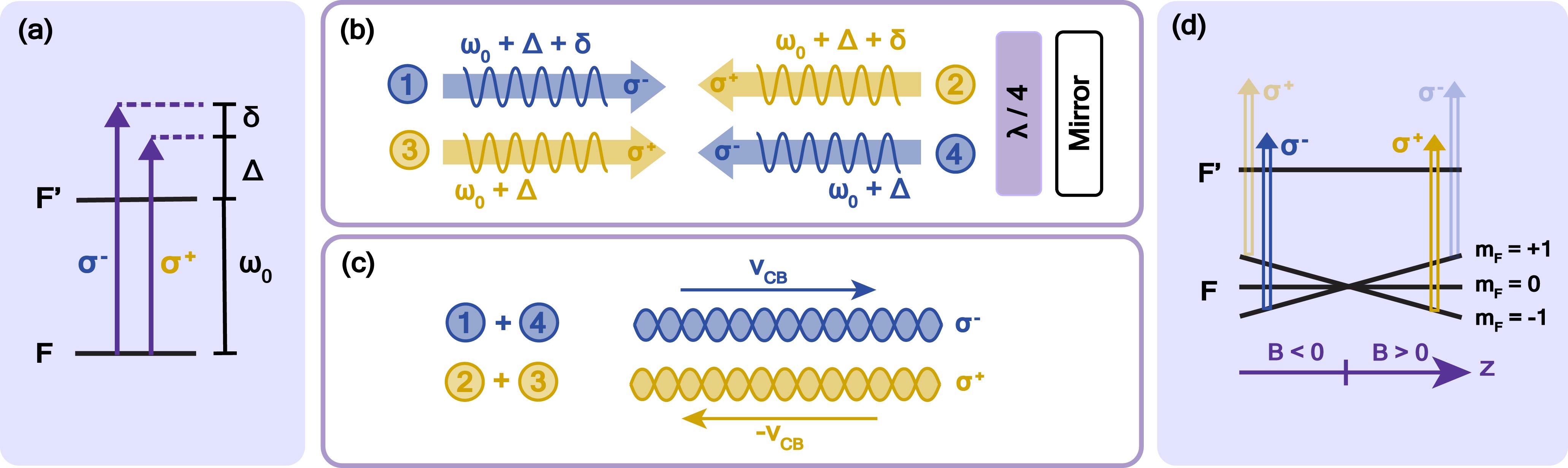}
    \caption{(a) Level diagram of the $F \to F' \leq F$ system, and the laser components. Lasers with $\sigma^+$ polarization are drawn in yellow, and lasers with $\sigma^-$ polarization are drawn in blue. (b) Frequency components of the MOT light in the $z$ direction. (c) The four frequency components add to two slowly-moving standing waves (conveyor belts) with velocities $\pm v_\text{CB}$. (d) Zeeman-shift due to the magnetic field gradient in the MOT leads to preferential interaction with one conveyor belt on each side of the MOT center. The conveyor belts are represented by the double arrows in this diagram.}
    \label{fig:mechanism}
\end{figure}

\medskip
In its simplest form, the scheme we propose uses two closely spaced laser frequencies ($\omega_\text{I}=\omega_0 + \Delta$ and $\omega_\text{II} = \omega_0 + \Delta + \delta$), with opposite circular polarizations (see Figure \ref{fig:mechanism}(a)). The MOT light comprises six laser beams directed along the \(\pm x\), \(\pm y\), and \(\pm z\) directions, with each pair of opposing beams having opposite polarizations, as shown in Figure \ref{fig:mechanism}(b).

\medskip
The light field that molecules experience is the sum of all six laser beams, each containing two frequency components. Taking the $z$ direction as an example, there are four laser components along $\pm z$, as shown in Fig.\ref{fig:mechanism}(b). The two counter-propagating $\sigma^-$ beams form a $\sigma^-$ standing wave. Due to the frequency splitting $\delta$, the standing wave moves with velocity $+v_\text{CB}= \frac{\delta}{2\omega_I}c$ (akin to a moving optical lattice). We shall refer to this as the ``$\sigma^-$ conveyor belt". Similarly, the $\sigma^+$ lasers form a $\sigma^+$ conveyor belt moving in the opposite direction with velocity $-v_\text{CB}$.

\medskip
A key to the induced trapping mechanism lies in the fact that, because of the magnetic field gradient in the MOT, a molecule preferentially interacts more strongly with one of the two conveyor belts depending on its position relative to the MOT center. If the magnetic field gradient in the $z$-direction is positive, and the ground state has $g > 0$, then a molecule at $z > 0$ is Zeeman-shifted closer to resonance with $\sigma^+$ light, and preferentially interacts with the $\sigma^+$ conveyor belt. In the frame of the conveyor belt, the molecule is moving with some velocity in a standing wave of $\sigma^+$ light. When the light is blue-detuned ($\Delta > 0$), the AC stark shift potential peaks have the highest optical pumping rate, leading to a Sisyphus-type cooling of the molecule in the frame of the $\sigma^+$ conveyor belt (similar to the cooling mechanism in \cite{emile1993magnetically}). In the lab frame, this means molecules get cooled into and pulled along by the $\sigma^+$ conveyor belt, which moves at a velocity $-v_\text{CB}$ towards the MOT center.

\medskip
Thus, if a molecule is located at \(z > 0\), it is picked up and carried by the \(\sigma^+\) conveyor belt, which moves with velocity \(v_\text{CB} < 0\). Conversely, if the molecule is at \(z < 0\), it is carried by the \(\sigma^-\) conveyor belt, which moves with velocity \(v_\text{CB} > 0\). In other words, molecules on either side of the MOT center are always directed towards the center by the corresponding conveyor belt. This mechanism, which ensures that molecules are consistently transported towards the MOT center (and then released near the center, see below), is what we refer to as the conveyor-belt trapping mechanism.

\medskip
The above description of molecular dynamics is for large Zeeman shifts, which occurs sufficiently away from the trap center. As the molecule riding on the conveyor belt approaches the center of the MOT, the magnitude of the magnetic field diminishes, causing the preference for one conveyor belt over the other to disappear. Consequently, a molecule is no longer bound to the conveyor belt in that region. Combined with the effect of regular gray molasses cooling \cite{devlin2016three} that occurs across the entire MOT region, the molecule slows down and remains near the MOT center.

\begin{figure}[ht]
    \centering
    \includegraphics[width=0.7\linewidth]{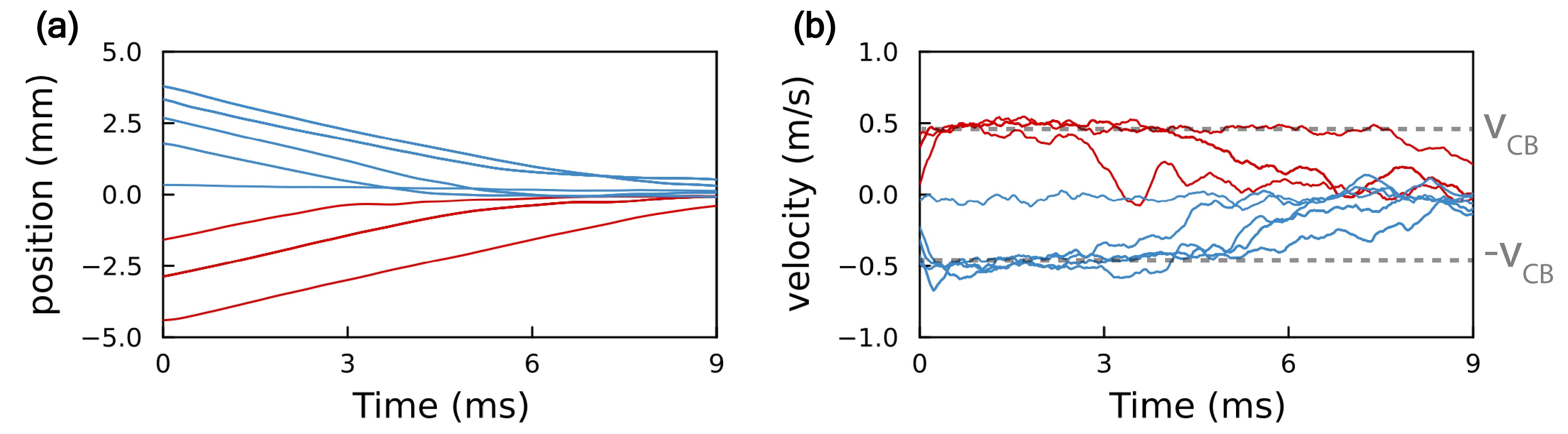}
    \caption{Trajectory and velocity in $z$-direction. Here we use a $F = 1 \to F' = 0$ system with transition frequency $\omega_0 = 479$ THz, transition linewidth $\Gamma = 6.4$ MHz, particle mass $m = 57$ amu, ground state $g$-factor 0.42, and  $\Delta = 10 \text{ MHz}, \delta = 1.5 \text{ MHz}$, saturation parameter $s = 5$ in each frequency component, and a magnetic field gradient of 10 G/cm. The trajectories are running-averaged for clarity of demonstration.} 
    \label{fig:trajectories}
\end{figure}

\medskip
To understand more thoroughly the proposed physical mechanism, we model a realistic blue-detuned conveyor-belt MOT using 3D Monte Carlo simulations based on stochastic Schrödinger equation methods \cite{dalibard1992wave}, see Figure \ref{fig:trajectories}. Particles are initially distributed randomly from a 1 mK thermal ensemble within a 1 mm diameter cloud. For clarity, particles starting at \(z > 0\) are colored blue while those starting at \(z < 0\) are colored red. The simulations reveal that particles on either side of the center are captured by the conveyor belt and quickly accelerated until they reach approximately \(\pm 0.5\) m/s = $v_\text{CB}$. Molecules maintain this velocity until they approach the MOT center, where they decelerate and are deposited near the center.

\section{Particle dynamics in the conveyor belts}
\label{dynamics}
The stochastic Schrödinger equations (SSE) simulation approach (used in the previous section) is found to be in excellent agreement with experiments \cite{CaOHBlueMOT}. However, more insight into dynamics can be gained from 1D acceleration curves obtained from optical Bloch equation (OBE) calculations. OBE methods were used for previous theoretical studies of molecular MOTs \cite{tarbuttOBE}, and found to be in good agreement with experiments. In this section, we calculate the acceleration experienced by molecules as a function of the magnetic field and velocity, and explore the molecular dynamics in the conveyor MOT.

\begin{figure}[ht]
    \centering
    \includegraphics[width=0.98 \linewidth]{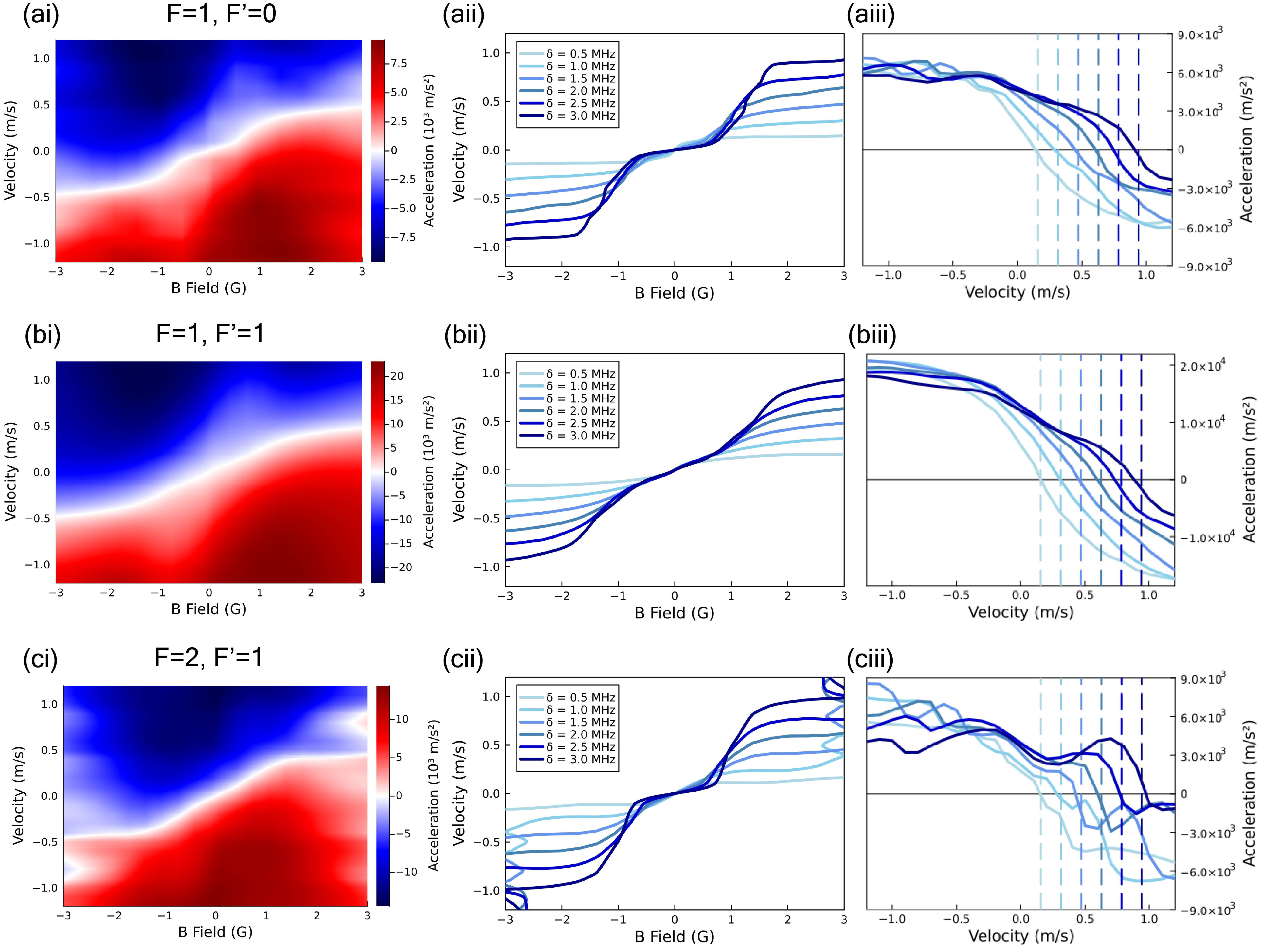}
    \caption{Acceleration function for a (a) $F = 1 \to F' = 0$, (b) $F = 1 \to F' = 1$, (c) $F = 2 \to F' = 1$ system with transition frequency $\omega_0 = 479$ THz, transition linewidth $\Gamma = 6.4$ MHz, particle mass $m = 57$ amu, ground state $g$-factor 0.42, and  $\Delta = 20 \text{ MHz}, \delta = 1.5 \text{ MHz}$, and $s$ = 5 in each component. (i) Acceleration function at $\delta$ = 1.5 MHz. (ii) The zero acceleration contour for force functions with $\delta = 0.5, 1.0, \dots, 3.0$ MHz. (iii) The acceleration vs. velocity curve at a fixed magnetic field of $2.5$ G.}
    \label{fig:smallDelta} 
\end{figure}

\medskip
Based on the conveyor-belt mechanism described in the previous section, we have certain expectations regarding the behavior of the acceleration as a function of magnetic field and velocity. Since trapping of molecules into the conveyor belt essentially functions as gray molasses cooling in the moving frame, we expect the molecules to have zero acceleration at \(v_\text{CB}\), which has opposite signs on either side of the MOT center. Additionally, the force should be positive when \(v < v_\text{CB}\) and negative when \(v > v_\text{CB}\), consistent with the characteristics of a cooling mechanism, as described above. These characteristics are indeed observed in our OBE model, as shown in Figure \ref{fig:smallDelta}(ai), (bi) and (ci), which depict the acceleration as a function of the magnetic field and velocity in a 2D heat map, in different type-II systems. The conveyor-belt velocity in this case is \(|v_\text{CB}| = 0.47 \, \text{m/s}\), which precisely matches the cooling features observed in regions where $|B| > 1.5 $ G. Figure \ref{fig:smallDelta}(aii), (bii) and (cii) show the zero acceleration contour for the same system, but with varying \(\delta\) values. Here the contour curves flatten at the corresponding conveyor-belt velocities, indicating the behavior expected from the physical intuition. Figure \ref{fig:smallDelta}(aiii), (biii) and (ciii) show a slice of the calculated acceleration at a fixed magnetic field of $B$ = 2.5 G. The intercept along the velocity axis aligns well with the corresponding conveyor-belt velocities, which are shown in dashed lines. The magnitude of the force is comparable in all these cases.

\begin{figure}[ht]
    \centering
    \includegraphics[width=0.7\linewidth]{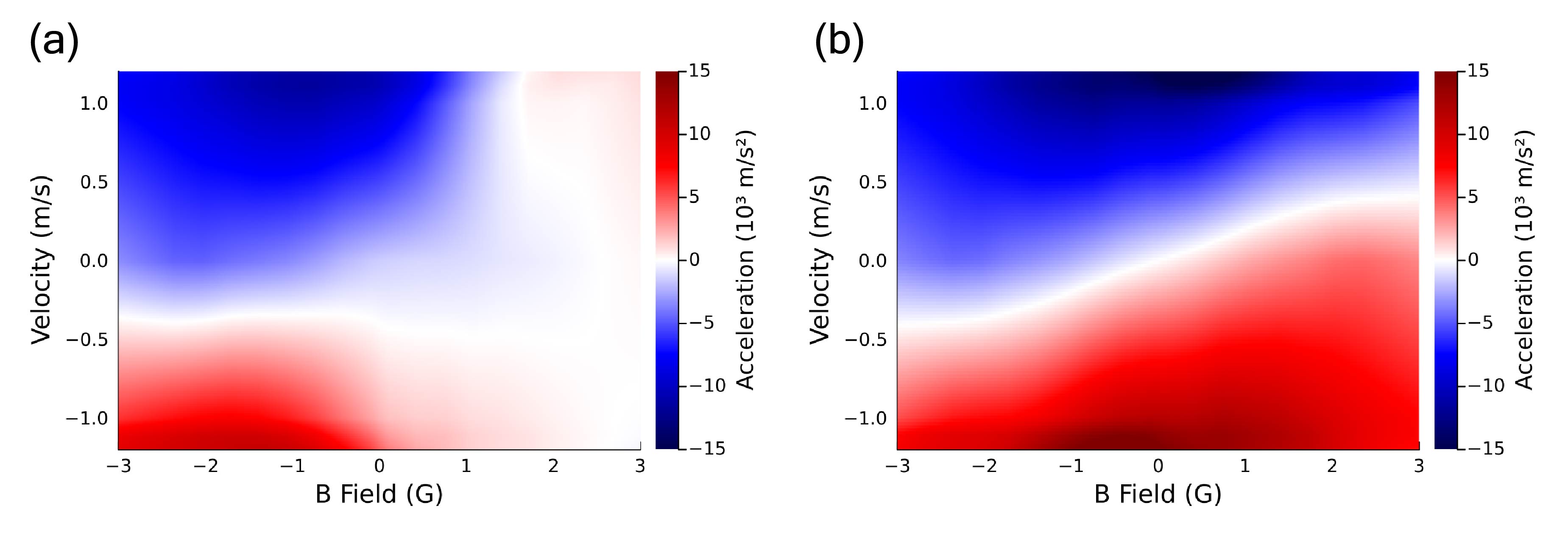}
    \caption{(a) The acceleration function with a single conveyor belt. (b) The acceleration function obtained by adding the acceleration from each conveyor belt.}
    \label{fig:oneBelt}
\end{figure}

\medskip
We consider the effect of a single conveyor belt by removing one of the conveyor belts in the $z$-direction. The resulting force function for a $F=1\to F'=1$ system is shown in Figure \ref{fig:oneBelt}(a). With only the \(\sigma^+\) conveyor belt present, the moving-frame cooling force is observed for \(B < 0\). The effect of the \(\sigma^+\) conveyor belt on the opposite side (\(B > 0\)) is more complex. Molecules at greater distances and higher $B$ fields experience a heating force from the \(\sigma^+\) conveyor belt, while particles near the center can still experience a cooling force. If we naively sum the forces from both conveyor belts, the resulting acceleration function, as shown in Figure \ref{fig:oneBelt}(b), reproduces the features from Figure \ref{fig:smallDelta}(c) reasonably well. This supports our physical interpretation that molecules essentially interact with one conveyor belt at a time, depending on which side of the trap the molecule is on.

\section{Characterization of the trapping force} \label{force curves}
In this section, to gain a deeper understanding of the conveyor MOT's performance, we examine how various characteristics, such as trap volume and capture velocity, depend on the conveyor MOT light parameters.

\begin{figure}[ht]
    \centering
    \includegraphics[width=0.95 \linewidth]{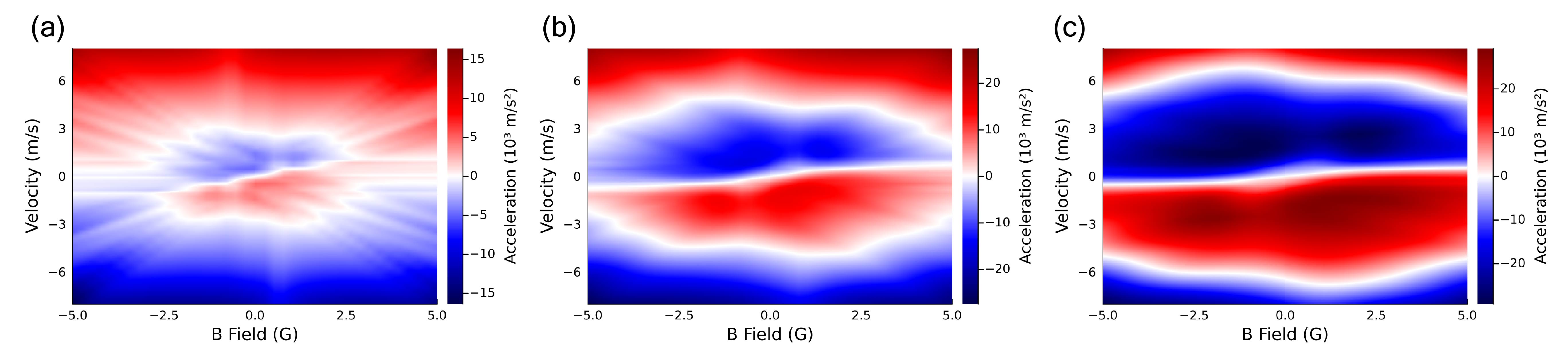}
    \caption{Acceleration functions for different laser powers for a $F=1\to F'=1$ system. The fixed laser parameters are $\Delta = 20$ MHz, $\delta$ = 1.5 MHz. The saturation of the lasers in each subplot is $s = 1, 3, 7$.}
    \label{fig:scanSaturation}
\end{figure}

\begin{figure}[ht]
    \centering
    \includegraphics[width=0.95\linewidth]{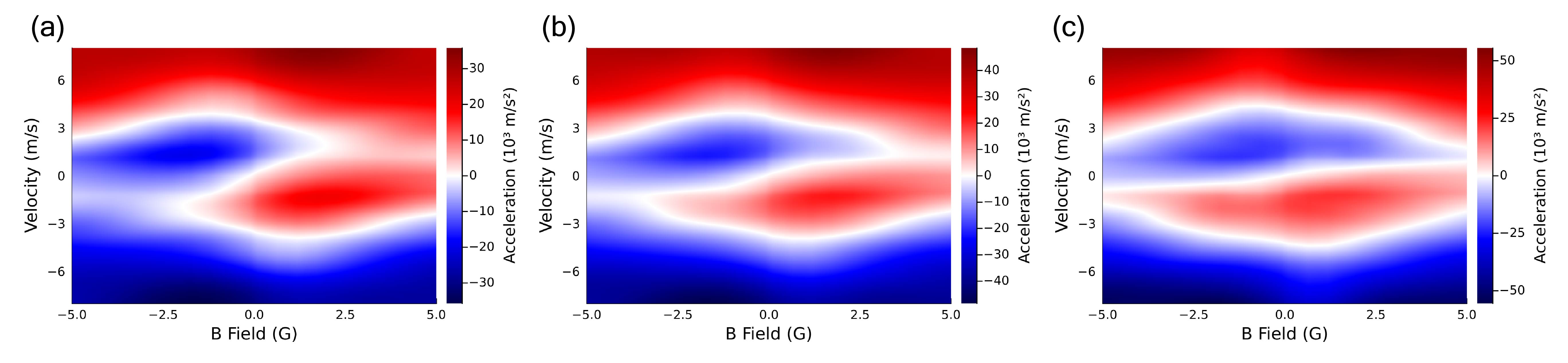}
    \caption{Acceleration functions for different laser detunings for a $F=1\to F'=1$ system. The fixed laser parameters are $s = 3$, $\delta$ = 1.5 MHz. The laser detuning in each subplot is $\Delta = 3, 5, 10$ MHz.}
    \label{fig:scanDelta}
\end{figure}

\medskip
The acceleration functions of an identical molecular system for different laser saturation parameters (\(s = 1, 3, 7\), where $s = I/I_\text{sat}$ and $I_\text{sat} = \frac{\pi h c \Gamma}{3 \lambda^3}$) are shown in Figure \ref{fig:scanSaturation}. The moving-frame cooling features observed in Figure \ref{fig:smallDelta} are evident in all three cases. Note the broader velocity range ($\pm$ 7 m/s) in Figure \ref{fig:scanSaturation} compared to Figure \ref{fig:smallDelta} ($\pm$ 1.2 m/s). This is done in order to observe the capture velocity characteristics of the MOT. Due to the blue detuning of the MOT light, Doppler heating at high velocities counteracts the gray molasses cooling in the conveyor belt frame, thereby limiting the MOT's capture velocity. From these plots, it is apparent that the capture velocity, which is indicated by the highest velocity at which a cooling force is present, increases with higher MOT light power. This is unsurprising since the conveyor MOT is essentially gray molasses cooling in two moving frames, and the relationship between capture velocity and laser parameters mirrors that observed in gray molasses and Sisyphus cooling \cite{devlin2016three}. The trap volume, which is indicated by the $B$ field where the moving-frame cooling feature disappears, also increases with laser power. The dependence on the one-photon laser detuning $\Delta$ is shown in Figure \ref{fig:scanDelta}, where $\Delta$ takes values of 3 MHz, 5 MHz and 10 MHz. The capture velocity remains roughly the same with different $\Delta$ values, whereas the trap volume increases with increased $\Delta$.

\section{Effect of excited-state Zeeman shift}
The systems we simulate so far all have zero excited state $g$-factor. While this clearly demonstrates the ground-state $g$-factor being the origin of the conveyor-belt trapping force, it raises the question of whether the conveyor-belt mechanism breaks down in the presence of a non-zero $g$-factor in the excited state. In this section, we address this question.

\begin{figure}[ht]
    \centering
    \includegraphics[width=0.65\linewidth]{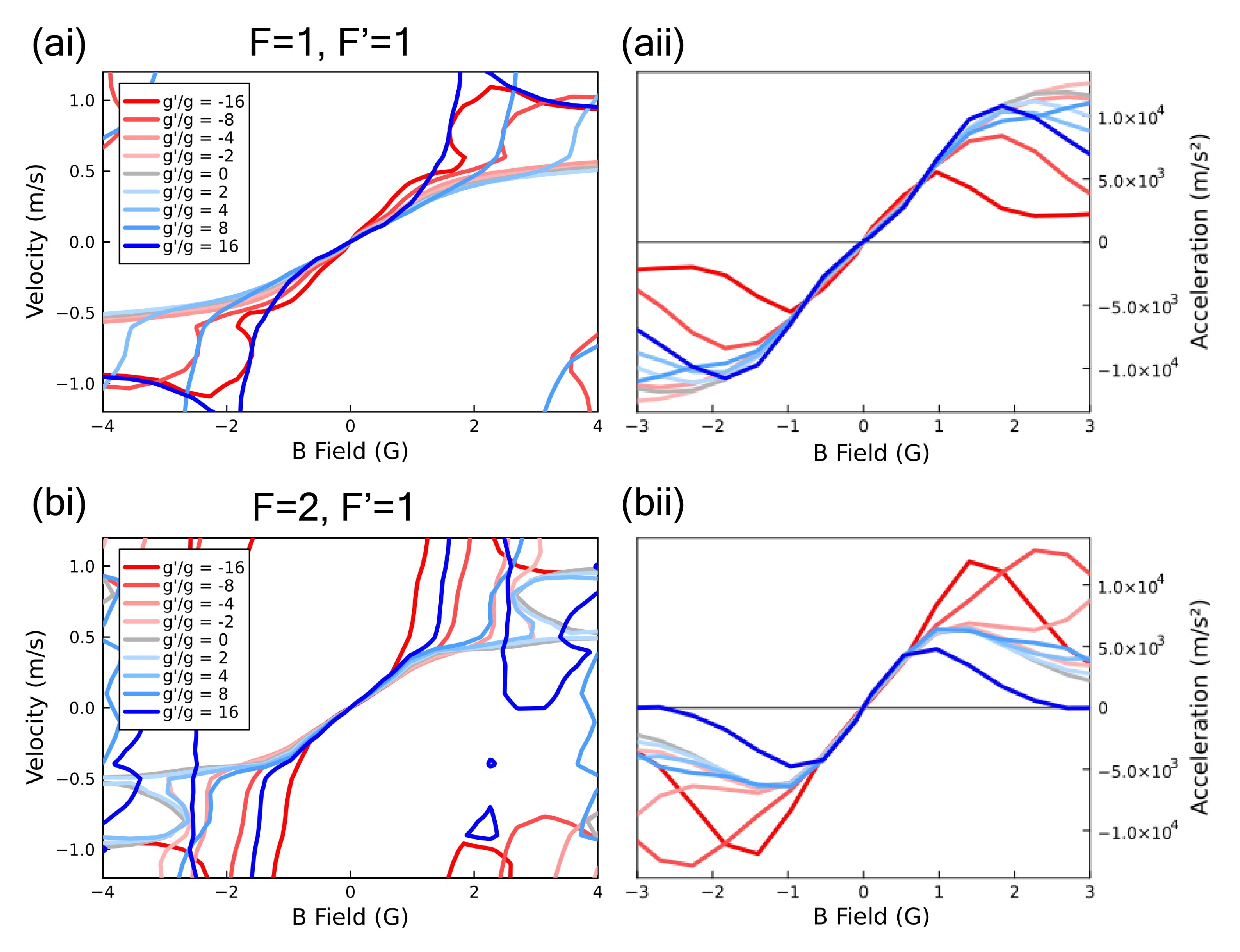}
    \caption{Effect of non-zero excited-state $g$-factors, in two different type-II systems: (a) $F=1\to F'=1$ and (b) $F=2 \to F'=1$. (ai), (bi) show the zero crossing of the force, and (aii), (bii) plot a slice of the acceleration function at zero velocity.}
    \label{fig:gfactors}
\end{figure}

Figure \ref{fig:gfactors} shows the zero force contours and a cross-section of the acceleration function for two different type-II systems: (a) $F = 1\to F'=1$ and (b) $F=2 \to F'=1$. ($F = 1 \to F'=0$ is excluded from this analysis since excited-state Zeeman shifts have no effect on this system.) Figure \ref{fig:gfactors}(ai) and (bi) show the zero crossing in the force function. In both cases, increasing the excited-state $g$-factor leads to a decreased trap volume, indicated by the signature plateau pattern of the conveyor-belt trapping vanishing at smaller $B$ fields. Figure \ref{fig:gfactors}(aii) and (bii) show a cross-section of the acceleration function at zero velocity. Near the center, the spring constant for these systems do not appear to change significantly with $g'$.

\section{Effect of complex structures}
In previous sections, we have described our studies of the conveyor belt mechanism in various type-II two-level systems. In this section, we investigate the force profile of conveyor-belt like laser configurations in more complex molecular systems, and confirm that the strong trapping provided by the conveyor-belt mechanism is still present in realistic molecular structures.

\begin{figure}[ht]
    \centering
    \includegraphics[width=0.95\linewidth]{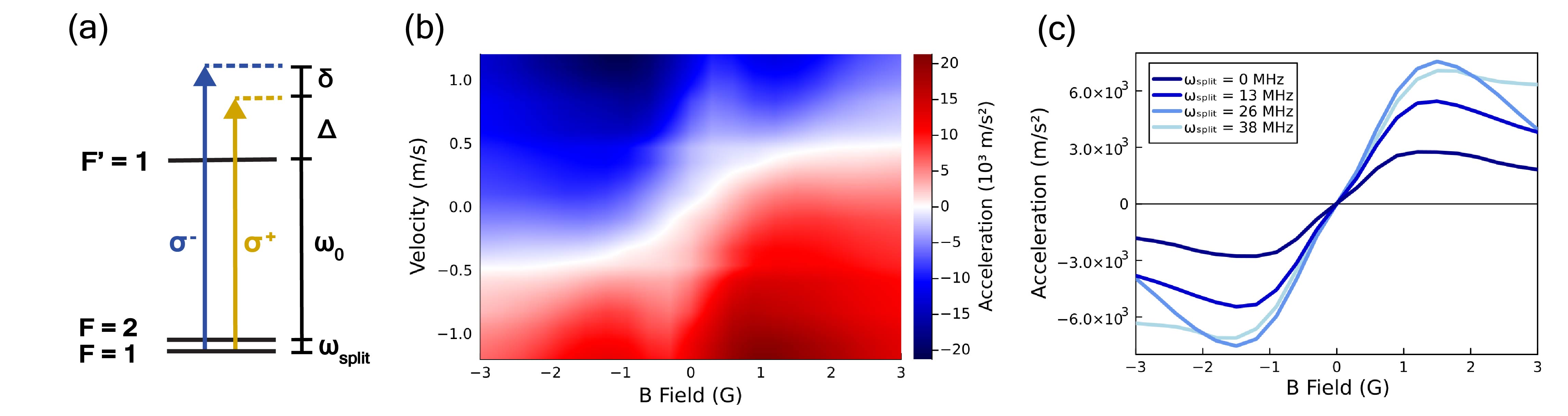}
    \caption{(a) The $F = 1,2\to F' = 1$ system, addressed by a pair of conveyor-belt lasers. (b) The acceleration function for this system with $\omega_\text{split}$ = 0, $\Delta$ = 20 MHz, $\delta$ = 1.5 MHz, and $s$ = 5.(c) The trapping force at zero velocity for different splittings $\omega_\text{split}$.}
    \label{fig:degenerate ground}
\end{figure}

\medskip
The excited state in the optical cycling transition usually has a non-zero electric dipole moment with multiple ground states. Here, we consider the effect of adding multiplicity to the \textit{ground states} of the simple $F \to F'$ system. There are two regimes here to consider: near-degenerate ground states (such as the hyperfine states in CaOH \cite{vilas2022magneto} and SrOH \cite{lasner2024magneto}), and well-resolved ground states (such as spin-rotation states, or hyperfine states in diatomic molecules such as CaF \cite{anderegg2017radio}, SrF \cite{barry2014magneto} and YO \cite{collopy20183d}). In the first case, since the splitting between the ground states is relatively small, both ground levels can be addressed by the same pair of conveyor belt lasers, as shown in Figure \ref{fig:degenerate ground}(a). Figure \ref{fig:degenerate ground}(b) shows the heatmap of the acceleration function for this system with $\omega_\text{split} = 0$, $\Delta$ = 20 MHz, $\delta$ = 1.5 MHz, $s$ = 5. The force profile turns out to be similar to that of the simple two-level system, with slightly decreased trapping force, which is expected since the ground state multiplicity has increased. Figure \ref{fig:degenerate ground}(c) shows a 1D slice of the force curve at zero velocity for different ground state splittings. As expected, the trapping force decreases as the splitting grows larger, since the $F=2 \to F'=1$ transition is moved further away from the resonance.

\begin{figure}[ht]
    \centering
    
    \includegraphics[width=0.75\linewidth]{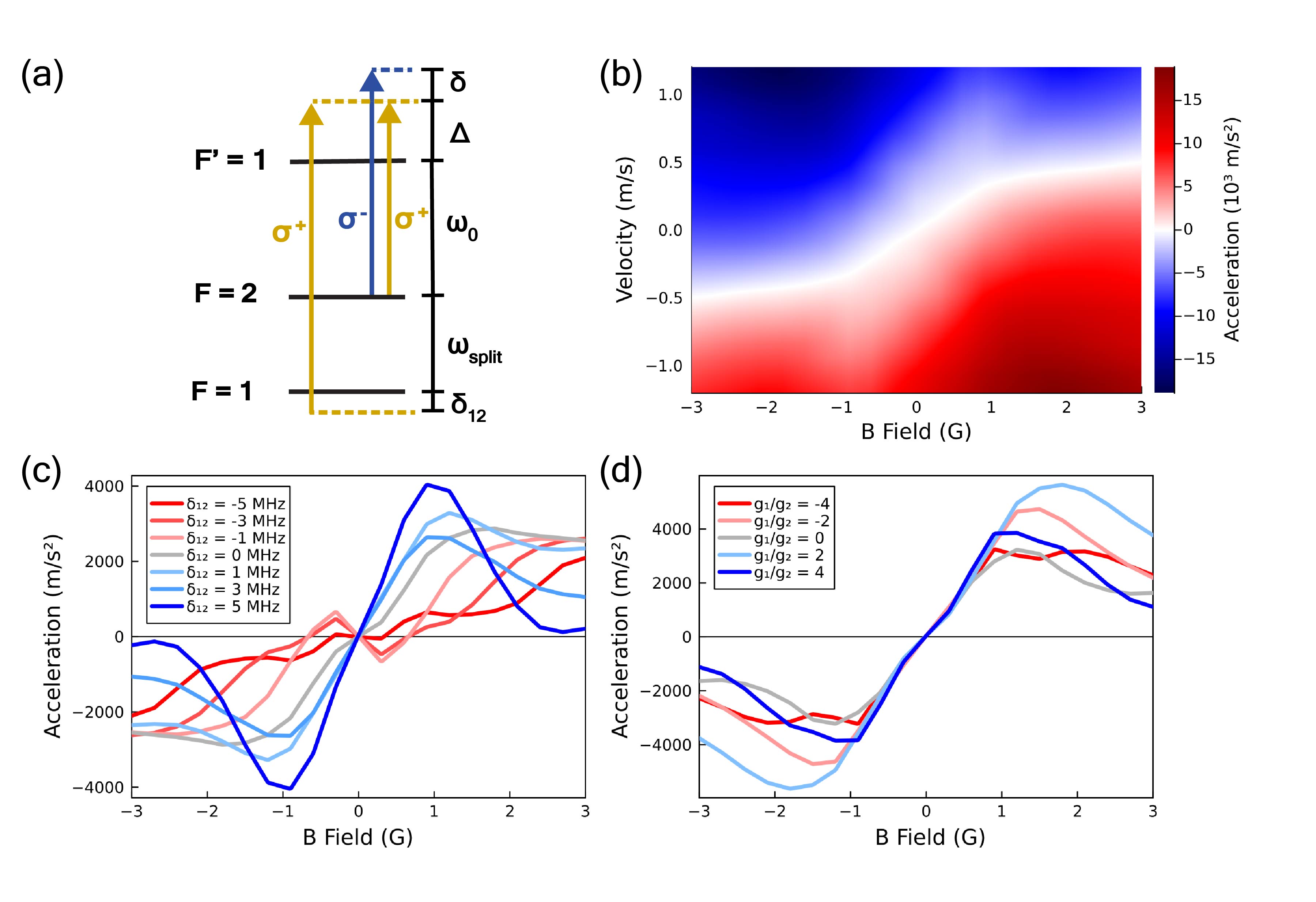}
     
     \caption{(a) The $F = 1,2\to F' = 1$ system, where the $F$ = 1 and $F$ = 2 states are far apart. The $F$ = 2 manifold is addressed by a pair of closely spaced conveyor belt lasers, whereas the $F = 1$ state is only addressed by one laser frequency. (b) The acceleration function for this system with $\omega_\text{split}$ = 100 MHz, $\Delta$ = 20 MHz, $\delta$ = 1.5 MHz, $\delta_{12} = 2$ MHz, $s$ = 5, and $g_1 = 0$. (c) The trapping force at zero velocity, for different two-photon detunings $\delta_{12}$ between the two pairs of conveyor-belt lasers. $g_1/g_2$ is fixed to 1. (d) The trapping force at zero velocity, for different $g$-factors in the $F=1$ ground state. $\delta_{12}$ is fixed at 2 MHz.}
    \label{fig:1_plus_2}
\end{figure}

We now consider the case where the two ground states are further apart than the radiative linewidth. In this case, they need to be addressed by separate laser components, which leads to further complications. For example, consider the scheme where a pair of conveyor belt lasers are used on one of the ground states, while the other ground state is addressed by only one laser component to prevent loss to dark states (Figure \ref{fig:1_plus_2}(a)). A heat map showing the acceleration $a(B,v)$ is shown for this case in Figure \ref{fig:1_plus_2}(b), and it looks similar to those we see in toy models in Sections \ref{dynamics} and \ref{force curves}. The choice of the third frequency also has a significant effect on the trapping force, as shown in Figure \ref{fig:1_plus_2}(c). Depending on the two-photon detuning $\delta_{12}$ between the $F=1$ laser frequency and the conveyor-belt lasers addressing $F=2$, the spring constant near the origin can even have different signs, consistent with experimental observations with CaOH in \cite{CaOHBlueMOT}. Figure \ref{fig:1_plus_2}(d) shows that the trapping is largely unaffected by the $g$-factor of the $F=1$ state.

\begin{figure}[ht]
    \centering
    
    \includegraphics[width=0.75\linewidth]{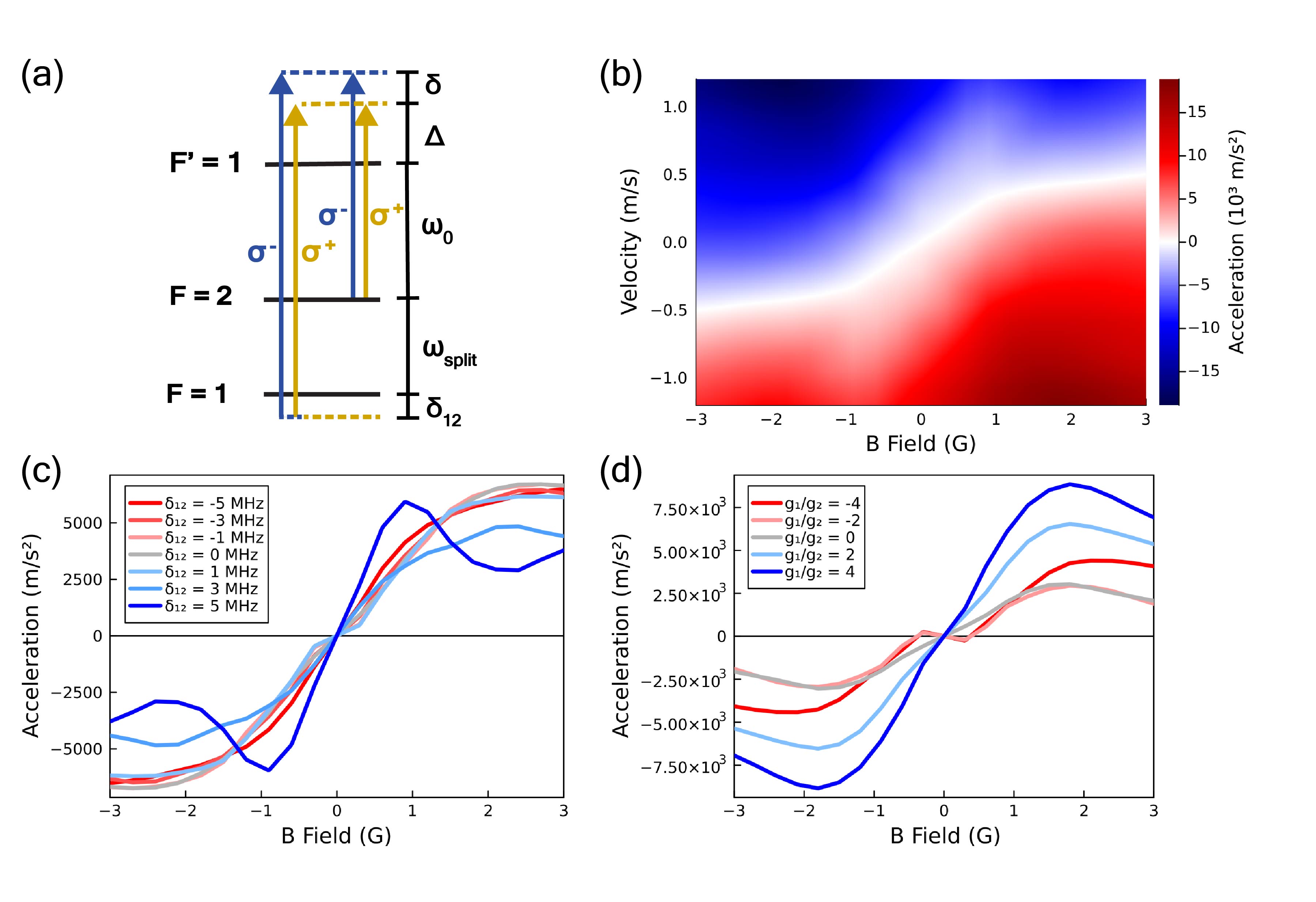}
    \caption{(a) The $F = 1,2\to F' = 1$ system, where the $F$ = 1 and $F$ = 2 states are far apart, addressed by two pairs of conveyor-belt lasers. (b) The acceleration function for this system with $\omega_\text{split}$ = 100 MHz, $\Delta$ = 20 MHz, $\delta$ = 1.5 MHz, $\delta_{12}$ = 0 MHz, $s$ = 5, and $g_1 =g_2=0.42$. (c) The trapping force at zero velocity, for different two-photon detunings $\delta_{12}$ between the two pairs of conveyor-belt lasers. $g_1/g_2$ is fixed to 1. (d) The trapping force at zero velocity, for different $g$-factors in the $F=1$ ground state. $\delta_{12}$ is fixed to 0 MHz.}
    \label{fig:2_plus_2}
\end{figure}

\medskip
In an alternative scheme, both ground states could be addressed by a \textit{pair} of conveyor belt lasers. The laser frequency components are shown in Figure \ref{fig:2_plus_2}(a). Figure \ref{fig:2_plus_2}(b) shows an example of the acceleration function of this system at zero two-photon detuning $\delta_{12}$ = 0. Again, this closely resembles the force profile of the simple toy systems described in Section \ref{dynamics}. In Figure \ref{fig:2_plus_2}(c) and (d), we study the trapping force at zero velocity at different two-photon detunings $\delta_{12}$ and different ground state $g$-factor ratios. In Figure \ref{fig:2_plus_2}(c), the acceleration (thus the spring constant) does not change significantly with $\delta_{12}$. In Figure \ref{fig:2_plus_2}(d), the spring constant is low for $g_1/g_2 < 0$, and increases with the ratio of $g_1/g_2$, which is expected since the pair of conveyor-belt lasers addressing $F=1$ provides anti-trapping if the $g$-factor of $F=1$ has the opposite sign as the $g$-factor of $F=2$.

\begin{figure}[ht]
    \centering
    \includegraphics[width=0.95\linewidth]{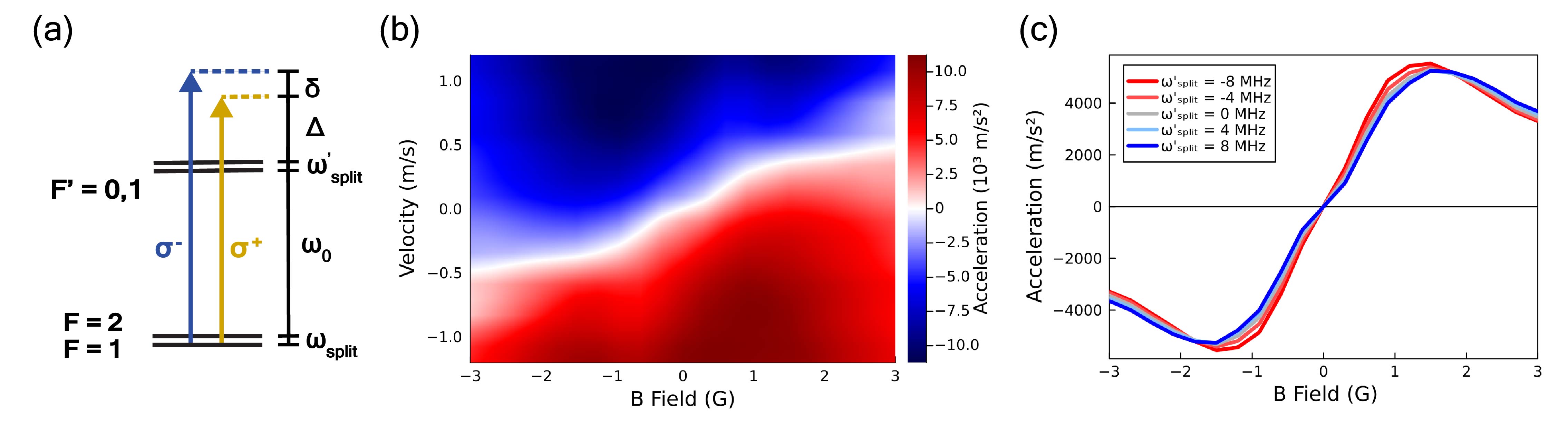}
     \caption{(a) The $F = 1,2\to F' = 1,0$ system, addressed by a pair of conveyor-belt lasers. (b) The acceleration function for this system with $\omega_\text{split}$ = 0, $\omega'_\text{split}$ = 0, $\Delta$ = 20 MHz, $\delta$ = 1.5 MHz, and $s$ = 5. (c) The trapping force at zero velocity, for different excited state splittings $\omega'_\text{split}$.}
    \label{fig:scan_excited}
\end{figure}

\medskip
Next, we consider the effect of degeneracy in the excited state. It is common for excited electronic states in molecules to have multiple unresolved or closely spaced hyperfine states. For example, the $J=1/2^-, F = 0, 1$ states in the excited A state of CaOH, CaF, SrF, SrOH all have sub-radiative linewidth level splitting. Figure \ref{fig:scan_excited}(b) shows the acceleration profile of the system given in Figure \ref{fig:scan_excited}(a), where both ground and excited states are degenerate. Here too, the force profile is essentially unchanged from the simple system in Section \ref{dynamics}. Figure \ref{fig:scan_excited}(c) scans the splitting between $F$ = 0 and $F$ = 1 levels in the excited state, and we see that the acceleration is unaffected, indicating that the conveyor-belt effect does not rely on near-degeneracy in the excited state.

\begin{figure}[ht]
    \centering
    \includegraphics[width=0.9\linewidth]{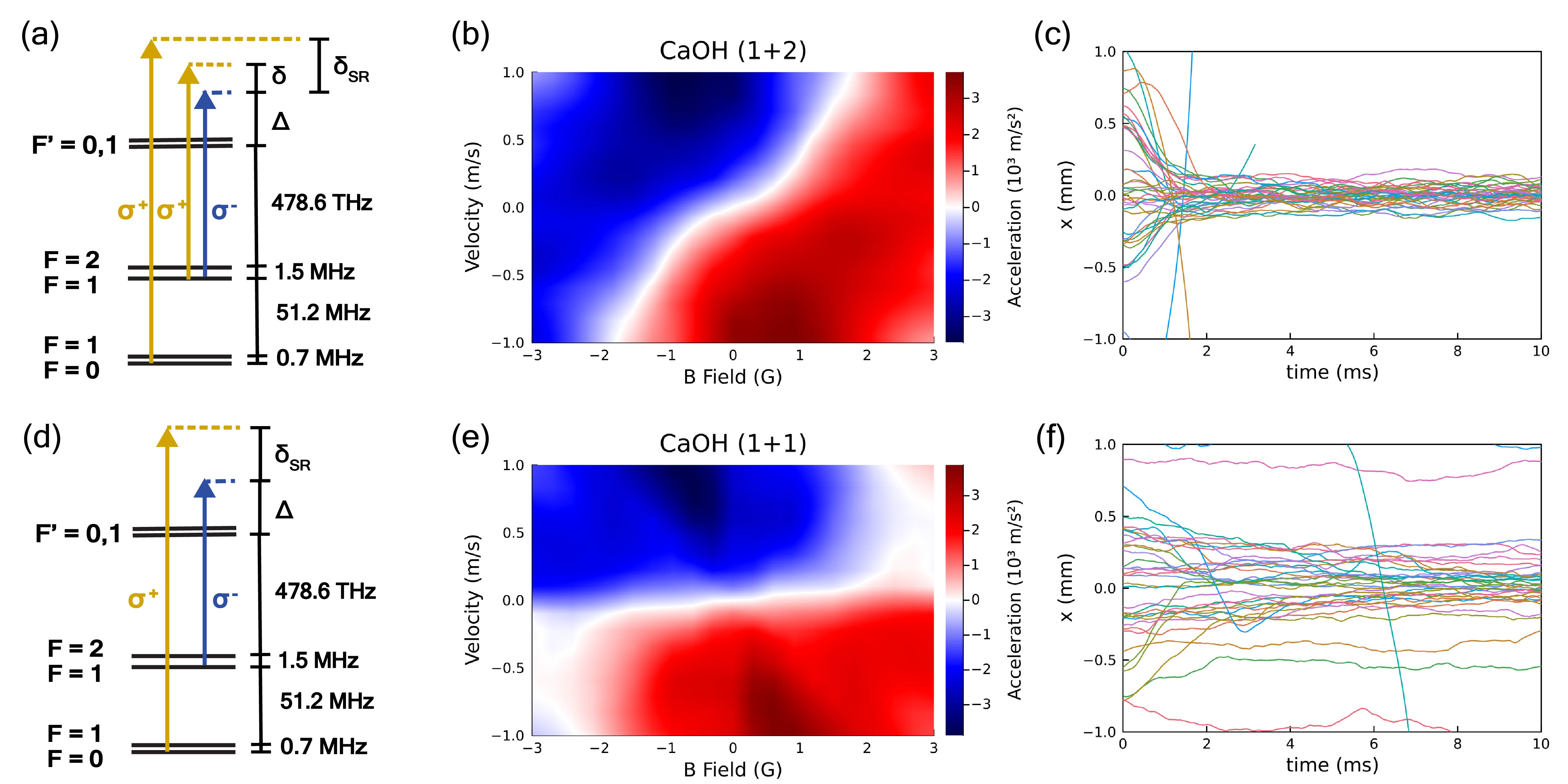}
    \caption{(a) Structure of the CaOH optical cycling transition, and the 1+2 configuration used for the conveyor-belt MOT in \cite{CaOHBlueMOT}. (b) The acceleration function for the 1+2 CaOH blue MOT. (c) The trajectories of molecules in the 1+2 MOT from SSE simulations.
    (d) The conventional 1+1 configuration for the CaOH blue MOT. (e) The acceleration function for the 1+1 CaOH blue MOT. (f) The trajectories of molecules in the 1+1 MOT from SSE simulations. The parameters used here are $\Delta$ = 5.1 MHz, $\delta$ = 1.75 MHz, $\delta_\text{SR}$ = 2.5 MHz, and $s$ = 1.5, 1.4, 1.1 for the three components in 1+2, and $s$ = 2 for both components in 1+1. The magnetic field gradient in the trajectory simulations is 30 G/cm.}
    \label{fig:CaOH}
\end{figure}

\begin{figure}[ht]
    \centering
    \includegraphics[width=0.9\linewidth]{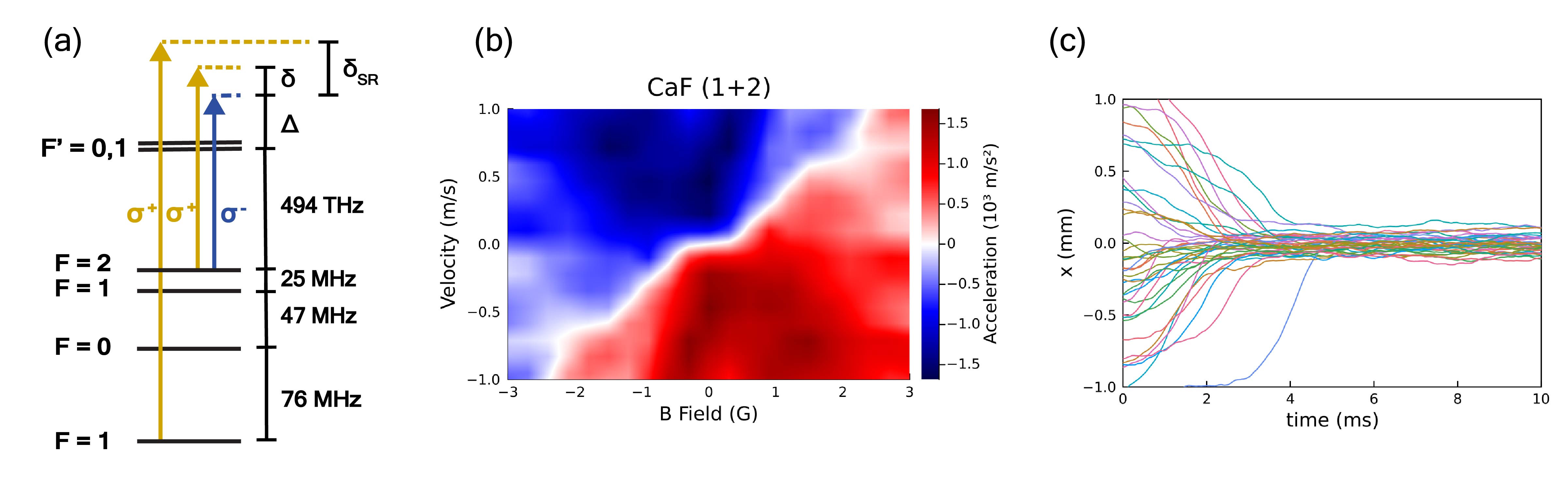}
    \caption{(a) Structure of the CaF optical cycling transition and the conveyor-belt configuration in \cite{CaFPaper}. (b) The acceleration function of the CaF conveyor MOT. (c) Trajectories of molecules in the conveyor MOT from SSE simulations. The parameters used here are $\Delta$ = 18.5 MHz, $\delta_\text{SR}$ = 1.5 MHz, $\delta$ = 2 MHz, and $s$ = 1.3 for all three components.}
    \label{fig:CaF}
\end{figure}

\medskip
Finally, we combine all the structural complexities discussed above, and demonstrate the conveyor-belt MOT in CaOH and CaF, using OBE and SSE analysis. In Figure \ref{fig:CaOH}, we compare the 1+2 conveyor-belt configuration (Figure \ref{fig:CaOH}(a)) and the conventional ``1+1" configuration (Figure \ref{fig:CaOH}(d)) for CaOH molecules, in which the $\sigma^+$ component comprising the conveyor-belt is absent. The acceleration functions of both are shown in Figure \ref{fig:CaOH}
(b,e), where we observe that the zero crossing is much steeper in the 1+2 case, which means molecules away from the center can accelerate to a higher velocity, indicating faster compression. Trajectories of molecules in both MOTs are simulated with SSE and shown in Figure \ref{fig:CaOH}(c,f). Here, we see that the 1+2 conveyor MOT has a much faster compression, and achieves a smaller cloud size compared to the 1+1 MOT, which is in agreement with experimental results in \cite{CaOHBlueMOT}. We also simulate the conveyor MOT for CaF molecules (Figure \ref{fig:CaF}), where the rapid and robust compression matches experimental observations as well \cite{CaFPaper}.

\section{Making a high capture velocity Conveyor MOT}
\begin{figure}[ht]
    \centering
    \includegraphics[width=0.8\linewidth]{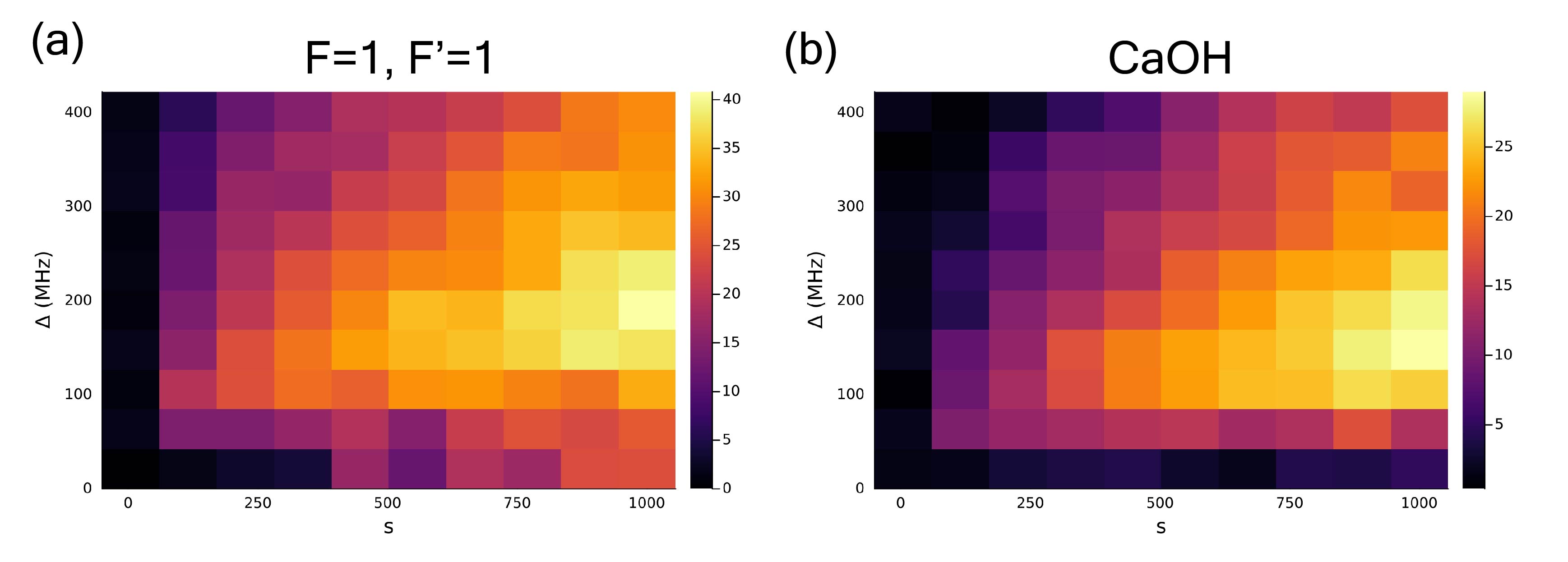}
    \caption{Capture velocity $v_c$ (m/s) of (a) a $F = 1 \to F' = 1$ system and (b) the CaOH molecule, as a function of laser detuning and saturation parameter. The MOT beams have a Gaussian profile with $1/e^2$ width of 1 cm.}
    \label{fig:highCapture}
\end{figure}

As discussed in Section \ref{force curves}, the capture velocity ($v_c$) and trap volume of the conveyor MOT increases with increased saturation parameter and detunings. By drastically increasing both $s$ and $\Delta$, it should be possible for a conveyor MOT to achieve a capture velocity that is comparable to, or higher than, a red-detuned MOT, as is now ubiquitously used to first trap molecules. In this capture scheme, because of the gray molasses cooling mechanism, once a molecule is slowed into the conveyor belt, its scattering rate will become and remain low, while it is being carried toward the MOT center. To quantify $v_c$, we again employ Monte Carlo simulations based on the SSE. By initializing molecules with a wide range (0 -- 50 m/s) of initial velocities, we calculate $v_c$ by evolving the system over a few milliseconds and identifying the trapped molecules with the highest initial velocity.\footnote{In the SSE simulation, the molecules are all initialized at the same height as the MOT center, and we observe that all molecules below $v_c$ are captured into the MOT. Realistically, molecules that are offset from the MOT height will have a lower capture velocity than $v_c$.} The dependence of $v_c$ on different detunings and laser powers is shown in Figure \ref{fig:highCapture}.

\medskip
An important feature of this scheme is that the slowing force is generated by the dipole force, which theoretically has no upper limit, unlike the radiative force, which is constrained by the maximum scattering rate and single photon recoil. By increasing laser power and detuning, $v_c$ of the conveyor-belt beams can be significantly increased. However, in practical scenarios, this capability will likely be limited by the onset of multi-photon processes at high power levels.

\medskip
Another advantage of the high capture velocity conveyor MOT is the relatively low scattering rate of the captured molecules. 
With the optimal parameters in both the simple $F=1 \to F'=1$ system and the CaOH molecule, SSE simulations show that the scattering rate drops to around $0.5\times10^6~\text{s}^{-1}$ (0.08$\Gamma$) once the molecules decelerate to the conveyor-belt velocity. This makes this configuration ``photon-efficient" when capturing from a continuous source---new incoming molecules can be slowed, while the already slowed molecules scatter few photons. When loading a red-detuned MOT from a molecular beam, the slowing and capture of incoming molecules relies on radiation pressure, where the force is generated by absorbing and scattering photons from a counter-propagating laser. 
However, because this force originates from light scattering, a problem in molecular MOTs is the limited photon budget due to decays into vibrational dark states. The far-detuned, high-intensity conveyor MOT that we propose in this section offers a possible circumvention of this issue.

\section{Summary and conclusions}
We propose and theoretically describe conveyor-belt trapping, which explains the enhanced trapping observed in recent experiments with CaOH  \cite{CaOHBlueMOT} and CaF \cite{CaFPaper}. Our numerical simulations, including both stochastic Schrödinger trajectory simulations and optical Bloch Equation force calculations, provide clear descriptions of the conveyor-belt mechanism at work. We investigate the trapping force, capture velocity, and trap volume across various types of systems, analyzing their dependence on laser detuning and power. Our results indicate that the conveyor-belt effect is ubiquitous in type-II systems, provided that the excited-state $g$-factor is not significantly higher than the ground-state $g$-factor. This suggests that a conveyor-belt MOT could be applicable to a wide range of molecular laser-cooling experiments, and extension to other applications may be possible.

\section{Acknowledgement}
The authors would like to acknowledge Yicheng Bao and Jiaqi You for valuable discussions in developing the numerical Monte-Carlo simulations. We also acknowledge the rest of the CaOH team (Lo\"{i}c Anderegg, Nathaniel Vilas, Paige Robichaud, Hana Lampson, Junheng Tao) and the CaF team (Scarlett Yu, Jiaqi You, Yicheng Bao, Dongkyu Lim) for making the experimental demonstrations \cite{CaOHBlueMOT, CaFPaper} possible, and for discussions and editing suggestions for the manuscript. This work was supported by the AFOSR, NSF, ARO, AOARD, DOE Quantum Systems Accelerator (QSA), and the CUA(PHY-2317134). GKL acknowledges support from the HQI.

\bibliographystyle{apsrev}
\bibliography{references}

\end{document}